\documentclass[twoside,twocolumn,english]{revtex4-1}
\usepackage[T1]{fontenc}
\usepackage[latin9]{inputenc}
\usepackage{geometry}
\usepackage{amsmath}
\usepackage{amssymb}
\usepackage{graphicx}

\usepackage{verbatim}

\usepackage{color,soul}

\usepackage{enumerate}

\usepackage{mathbbol}
\usepackage{amsfonts}
\usepackage{ulem}
\usepackage{float}

\makeatletter
\RequirePackage{colortbl, tabularx}
\@ifundefined{comment}{}
  {
   
  }%

\makeatother

\usepackage{babel}

\setstcolor{red}

\begin{document}
\global\long\def\l{\lambda}%
\global\long\def\ints{\mathbb{Z}}%
\global\long\def\nat{\mathbb{N}}%
\global\long\def\re{\mathbb{R}}%
\global\long\def\com{\mathbb{C}}%
\global\long\def\dff{\triangleq}%
\global\long\def\df{\coloneqq}%
\global\long\def\del{\nabla}%
\global\long\def\cross{\times}%
\global\long\def\der#1#2{\frac{d#1}{d#2}}%
\global\long\def\bra#1{\left\langle #1\right|}%
\global\long\def\ket#1{\left|#1\right\rangle }%
\global\long\def\braket#1#2{\left\langle #1|#2\right\rangle }%
\global\long\def\ketbra#1#2{\left|#1\right\rangle \left\langle #2\right|}%
\global\long\def\paulix{\begin{pmatrix}0  &  1\\
 1  &  0 
\end{pmatrix}}%
\global\long\def\pauliy{\begin{pmatrix}0  &  -i\\
 i  &  0 
\end{pmatrix}}%
\global\long\def\sinc{\mbox{sinc}}%
\global\long\def\ft{\mathcal{F}}%
\global\long\def\dg{\dagger}%
\global\long\def\bs#1{\boldsymbol{#1}}%
\global\long\def\norm#1{\left\Vert #1\right\Vert }%
\global\long\def\H{\mathcal{H}}%
\global\long\def\tens{\varotimes}%
\global\long\def\rationals{\mathbb{Q}}%
 
\global\long\def\tri{\triangle}%
\global\long\def\lap{\triangle}%
\global\long\def\e{\varepsilon}%
\global\long\def\broket#1#2#3{\bra{#1}#2\ket{#3}}%
\global\long\def\dv{\del\cdot}%
\global\long\def\eps{\epsilon}%
\global\long\def\rot{\vec{\del}\cross}%
\global\long\def\pd#1#2{\frac{\partial#1}{\partial#2}}%
\global\long\def\L{\mathcal{L}}%
\global\long\def\inf{\infty}%
\global\long\def\d{\delta}%
\global\long\def\I{\mathbb{I}}%
\global\long\def\D{\Delta}%
\global\long\def\r{\rho}%
\global\long\def\hb{\hbar}%
\global\long\def\s{\sigma}%
\global\long\def\t{\tau}%
\global\long\def\O{\Omega}%
\global\long\def\a{\alpha}%
\global\long\def\b{\beta}%
\global\long\def\th{\theta}%
\global\long\def\l{\lambda}%

\global\long\def\Z{\mathcal{Z}}%
\global\long\def\z{\zeta}%
\global\long\def\ord#1{\mathcal{O}\left(#1\right)}%
\global\long\def\ua{\uparrow}%
\global\long\def\da{\downarrow}%
 
\global\long\def\co#1{\left[#1\right)}%
\global\long\def\oc#1{\left(#1\right]}%
\global\long\def\tr{\mbox{tr}}%
\global\long\def\o{\omega}%
\global\long\def\nab{\del}%
\global\long\def\p{\psi}%
\global\long\def\pro{\propto}%
\global\long\def\vf{\varphi}%
\global\long\def\f{\phi}%
\global\long\def\mark#1#2{\underset{#2}{\underbrace{#1}}}%
\global\long\def\markup#1#2{\overset{#2}{\overbrace{#1}}}%
\global\long\def\ra{\rightarrow}%
\global\long\def\cd{\cdot}%
\global\long\def\v#1{\vec{#1}}%
\global\long\def\fd#1#2{\frac{\d#1}{\d#2}}%
\global\long\def\P{\Psi}%
\global\long\def\dem{\overset{\mbox{!}}{=}}%
\global\long\def\Lam{\Lambda}%
 
\global\long\def\m{\mu}%
\global\long\def\n{\nu}%

\global\long\def\ul#1{\underline{#1}}%
\global\long\def\at#1#2{\biggl|_{#1}^{#2}}%
\global\long\def\lra{\leftrightarrow}%
\global\long\def\var{\mbox{var}}%
\global\long\def\E{\mathcal{E}}%
\global\long\def\Op#1#2#3#4#5{#1_{#4#5}^{#2#3}}%
\global\long\def\up#1#2{\overset{#2}{#1}}%
\global\long\def\down#1#2{\underset{#2}{#1}}%
\global\long\def\lb{\biggl[}%
\global\long\def\rb{\biggl]}%
\global\long\def\RG{\mathfrak{R}_{b}}%
\global\long\def\g{\gamma}%
\global\long\def\Ra{\Rightarrow}%
\global\long\def\x{\xi}%
\global\long\def\c{\chi}%
\global\long\def\res{\mbox{Res}}%
\global\long\def\dif{\mathbf{d}}%
\global\long\def\dd{\mathbf{d}}%
\global\long\def\grad{\vec{\del}}%

\global\long\def\mat#1#2#3#4{\left(\begin{array}{cc}
#1 & #2\\
#3 & #4
\end{array}\right)}%
\global\long\def\col#1#2{\left(\begin{array}{c}
#1\\
#2
\end{array}\right)}%
\global\long\def\sl#1{\cancel{#1}}%
\global\long\def\row#1#2{\left(\begin{array}{cc}
#1 & ,#2\end{array}\right)}%
\global\long\def\roww#1#2#3{\left(\begin{array}{ccc}
#1 & ,#2 & ,#3\end{array}\right)}%
\global\long\def\rowww#1#2#3#4{\left(\begin{array}{cccc}
#1 & ,#2 & ,#3 & ,#4\end{array}\right)}%
\global\long\def\matt#1#2#3#4#5#6#7#8#9{\left(\begin{array}{ccc}
#1 & #2 & #3\\
#4 & #5 & #6\\
#7 & #8 & #9
\end{array}\right)}%
\global\long\def\su{\uparrow}%
\global\long\def\sd{\downarrow}%
\global\long\def\coll#1#2#3{\left(\begin{array}{c}
#1\\
#2\\
#3
\end{array}\right)}%
\global\long\def\h#1{\hat{#1}}%
\global\long\def\colll#1#2#3#4{\left(\begin{array}{c}
#1\\
#2\\
#3\\
#4
\end{array}\right)}%
\global\long\def\check{\checked}%
\global\long\def\v#1{\vec{#1}}%
\global\long\def\S{\Sigma}%
\global\long\def\F{\Phi}%
\global\long\def\M{\mathcal{M}}%
\global\long\def\G{\Gamma}%
\global\long\def\im{\mbox{Im}}%
\global\long\def\til#1{\tilde{#1}}%
\global\long\def\kb{k_{B}}%
\global\long\def\k{\kappa}%
\global\long\def\ph{\phi}%
\global\long\def\el{\ell}%
\global\long\def\en{\mathcal{N}}%
\global\long\def\asy{\cong}%
\global\long\def\sbl{\biggl[}%
\global\long\def\sbr{\biggl]}%
\global\long\def\cbl{\biggl\{}%
\global\long\def\cbr{\biggl\}}%
\global\long\def\hg#1#2{\mbox{ }_{#1}F_{#2}}%
\global\long\def\J{\mathcal{J}}%
\global\long\def\diag#1{\mbox{diag}\left[#1\right]}%
\global\long\def\sign#1{\mbox{sgn}\left[#1\right]}%
\global\long\def\T{\th}%
\global\long\def\rp{\reals^{+}}%

\title{Diffusion with Local Resetting and Exclusion}
\author{Asaf Miron$^{1}$ and Shlomi Reuveni$^{2}$}
\address{$\mbox{ }^{1}$Department of Physics of Complex Systems, Weizmann
Institute of Science, Rehovot 7610001, Israel}
\address{$\mbox{ }^{2}$School of Chemistry, The Center for Physics and Chemistry
of Living Systems, \& The Mark Ratner Institute for Single Molecule
Chemistry, \& The Raymond and Beverly Sackler Center for Computational
Molecular and Materials Science, Tel Aviv University, Tel Aviv 6997801,
Israel}

\begin{abstract}
Stochastic resetting models diverse phenomena across numerous scientific disciplines. Current understanding stems from the renewal framework, which relates systems subject to \textit{global} resetting to their non-resetting counterparts. Yet, in interacting many-body systems, even the simplest scenarios involving resetting give rise to the notion of \textit{local} resetting, whose analysis falls outside the scope of the renewal approach. A prime example is that of diffusing particles with excluded volume interactions that \textit{independently} attempt to reset their position to the origin of a 1D lattice. With renewal rendered ineffective, we instead employ a mean-field approach whose validity is corroborated via extensive numerical simulations. The emerging picture sheds first light on the non-trivial interplay between interactions and resetting in many-body systems.
\end{abstract}

\maketitle

As allegorized in the popular nursery rhyme on a small spider's Sisyphean attempt to ascend a waterspout during questionable weather, surmounting a long and volatile endeavor is often quite challenging and full of setbacks. Similar setbacks also arise in many natural phenomena that are prone to ``resetting'': the abrupt cessation of a dynamical process, which consecutively starts anew. Examples for resetting phenomena are abundant and come from a diverse range of research fields including search, first-passage, and animal foraging \cite{Search1,Search2, Search3,Search4,Search5,Search6,Search7,Search8,Search9,Search10}, algorithmics \cite{Luby, Gomes, Steiger}, optimization \cite{Optimization1,Optimization2,Optimization3,Optimization4,Optimization5}, and reaction kinetics \cite{Reuveni, Berezhkovskii, Robin, Lapeyre, Roldan1, Budnar}.

The renewal framework constitutes the current state-of-the-art in the analysis of resetting phenomena \cite{Search6,Search7,renewalf1,renewalf1}. This powerful approach utilizes the understanding that a resetting event forces the system back to its initial state to formulate stochastic renewal equations, relating the properties of resetting systems to those of their non-resetting counterparts. Such relations have proven salient in uncovering  steady-state, transport, relaxation, and first-passage properties in a long list of model systems \cite{renewal1,renewal2,renewal3,renewal4,renewal5,renewal6,renewal7,renewal8,renewal9,renewal10,renewal11,renewal12,renewal13,renewal14,renewal15,renewal16,renewal17,renewal18,renewal19,renewal20}. Moreover, they have been instrumental in the discovery of universal phenomena that emerge as a result of resetting \cite{Search6,Search7,Search9,Optimization2,universal1,universal2,universal3,universal4,universal5}.  

\begin{figure}[t]
\begin{centering}
\includegraphics[scale=0.23]{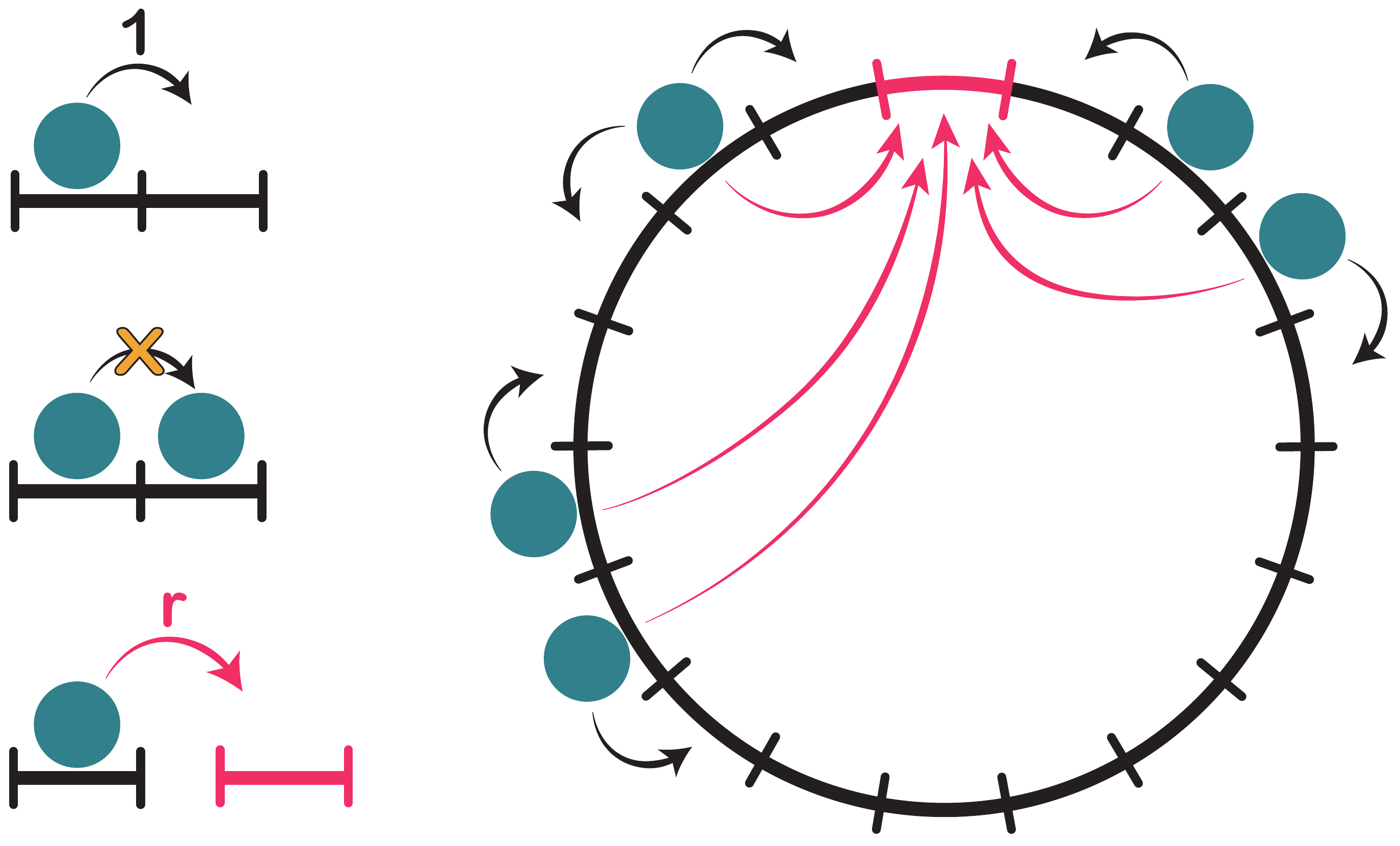}
\par\end{centering}
\caption{An illustration of diffusion with exclusion and local resetting on a ring lattice. Each particle \textit{attempts} to hop to its left and right neighboring sites with rate $1$, and to reset its position to the origin with rate $r$. In both cases, an attempt is successful only if the target site is vacant.} \label{illustration}
\end{figure}

In spite of its centrality to the study of resetting phenomena, the renewal framework's applicability is limited. Specifically, in many-body systems, it only applies  when resetting acts to bring the \textit{entire} system back to its initial configuration, a process which we hereby term ``global resetting'' \cite{global1,global2,global3}. Yet, generically, resetting will have a more local nature that can, in turn, be affected by interactions. As a concrete example, consider diffusive particles, subject to repulsive short-ranged interactions, that move along a 1D track immersed in a fluid. In this case, resetting can be used to model the unbinding of a particle from the track and its subsequent rebinding at a uniquely favorable location. However, due to interactions, rebinding cannot occur if this location is already occupied by a different particle. This, in fact, is precisely the picture that arises in bio-polymerization, where RNA polymerases and ribosomes play the role of resetting particles. Since a resetting event leaves the system's configuration mostly intact, aside from the new position of the resetting particle, the renewal framework does not apply and alternative approaches must be sought. 
The absence of a unifying theoretical framework renders the study of ``local'' resetting in interacting many-body systems extremely challenging. 

In this Letter, we take a first step towards establishing an understanding of local resetting within an analytically tractable model. In particular, we study a 1D ring lattice of $L$ sites, occupied by diffusive particles which interact via volume exclusion and \textit{independently} attempt to reset their position to the origin site, if it is vacant (see Fig. \ref{illustration}). These dynamics give rise to a non-trivial steady-state density profile, whose analysis lies well beyond the scope of renewal theory. Instead, taking a mean-field (MF) approach, we are able to derive a closed-form solution for the stationary density profile and analyze its scaling properties in the limit of large $L$. The MF description is corroborated against extensive numerical simulations to remarkable accuracy. The results established herein pave the way to the extension of current experimental studies of single-particle resetting \cite{Roichman2020,Ciliberto2020}, to interacting many-body systems.

\textit{The Model} - Consider a $1D$ periodic lattice of $L$ sites
labeled $\el=0,...,L-1,$ occupied by $N$ particles of average density
$\overline{\r}\equiv N/L$. The particles are subject to hard-core,
exclusion interactions by which each site may hold one particle at
most \citep{harris1965diffusion,jepsen1965dynamics,percus1974anomalous,alexander1978diffusion}. The system then evolves in continuous time via the dynamical rules illustrated in Fig. \ref{illustration}: Each particle \textit{attempts} to hop to its left
and right neighboring sites with rate $1$ and to reset its position
to site $\el=0$ with rate $r$. In both cases, an attempt is successful
only if the target site is vacant. 

These dynamics can be formulated in terms of a Markov chain. Let $\t_{\el}\left(t\right)$
denote the occupation of site $\el$ at time $t$, taking the value
$0$ if the site is vacant and $1$ otherwise. The model's dynamics
imply the following evolution of $\t_{\el}\left(t\right)$
\begin{equation}
\t_{\el}\left(t+dt\right)-\t_{\el}\left(t\right)=\G_{\el}\left(t\right),\label{eq:markov chain}
\end{equation}
where $\G_{\el}\left(t\right)$ is given by
\begin{equation}
\G_{\el}\left(t\right)=\begin{cases}
\t_{\el\pm1} & \text{w.p. }\left(1-\t_{\el}\right)dt\\
-\t_{\el} & \text{w.p. }\left(1-\t_{\el\pm1}\right)dt\\
\s_\el & \text{w.p. }\left(1-\t_{0}\right)R_{\el}dt
\end{cases}\label{eq:markov Gamma}
\end{equation}
and w.p. abbreviates ``with probability''. Clearly, a distinction
must be made between site $\el=0$, which experiences an influx of
resetting particles, and the remaining sites $\el\ne0$. Equation \eqref{eq:markov chain} for $\G_{\el}\left(t\right)$ indeed serves as short-hand notation for  
\begin{equation}
\begin{array}{c}
\s_\el=-1\text{ and }R_{\el}=r\t_{\el}\,\,\,\,\,\,\,\,\,\,\,\,\,\,\,\,\,\,\,\,\,\,\,\text{ for }\el\ne0\\
\s_\el=+1\text{ and }R_{\el}=r\sum_{m=1}^{L-1}\t_{m}\,\,\text{  for }\el=0
\end{array}.\label{eq:Gamma_sigma_R}
\end{equation}
Note that the expression for $R_{0}$ is not arbitrary, originating from the model's particle conservation, i.e. $\sum_{\el=0}^{L-1}\left[\t_{\el}\left(t+dt\right)-\t_{\el}\left(t\right)\right]=0$. 

\textit{Main Results} - Using an analytical mean-field approach and extensive numerical simulations, we show that the stationary density profile of resetting and interacting random walkers behaves very differently from that found in the absence of interactions. As in the non-interacting case, resetting acts to concentrate particles at the origin site $\el=0$. However, here, exclusion prevents the origin from being occupied by more than a single particle. Moreover, resetting cannot occur when the origin is occupied, giving rise to a non-trivial interplay between exclusion, diffusion, and resetting.

\begin{figure}
\begin{centering}
\includegraphics[scale=0.6]{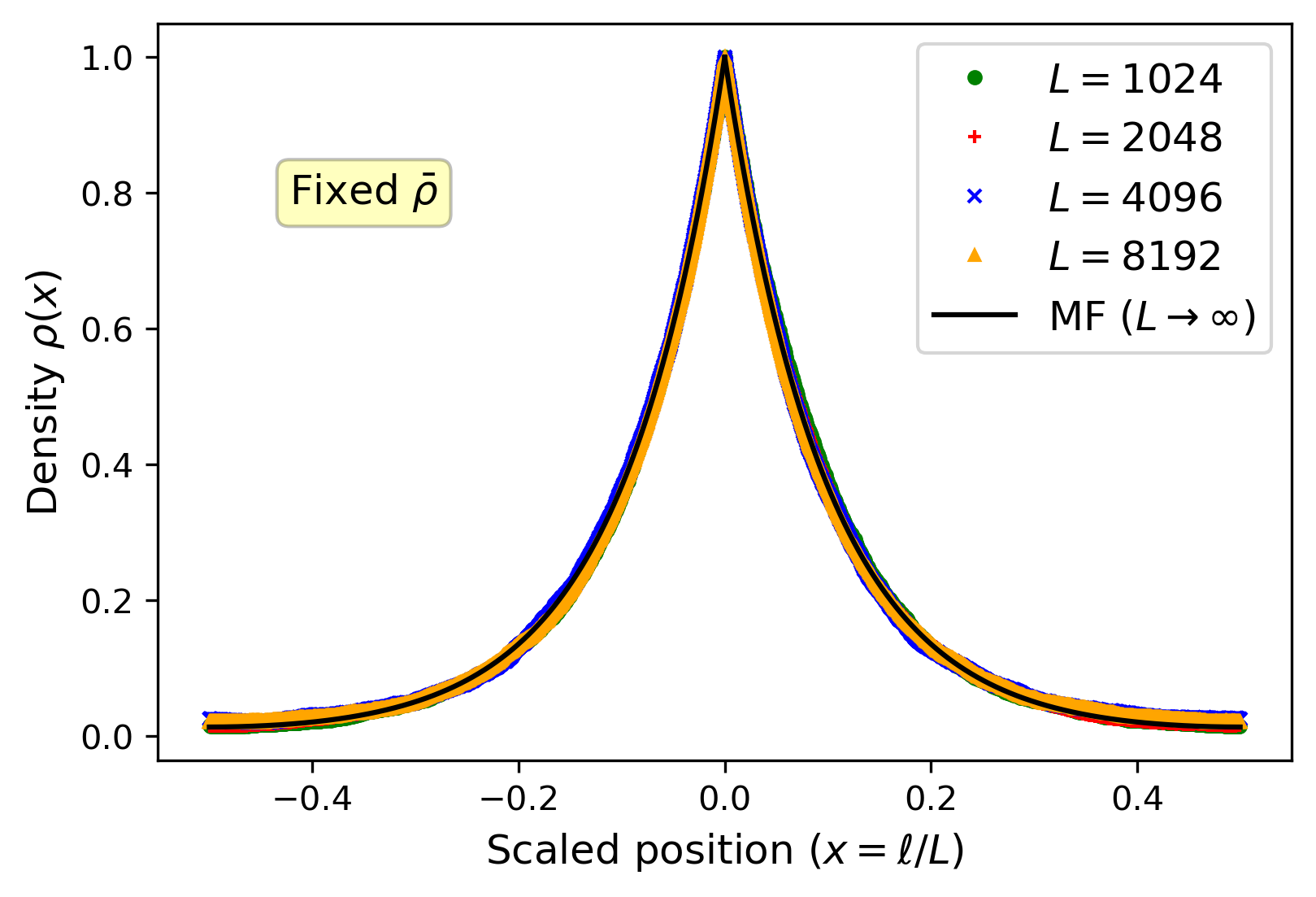}
\par\end{centering}
\caption{Data collapse of the density profile for $r=1$ and $\overline{\protect\r}=0.2$
versus the scaling variable $x=\protect\el/L$. The different markers
denote various values of the system size $L$ while the solid black
curve denotes the theoretical MF prediction in Eq. \eqref{eq:rho_bulk_sol_3} for $L\ra\infty$. Note that the site indices $\el$ have been shifted to $\el=-L/2+1,...,0,...,L/2$ for convenience of presentation. Correspondingly, $x\in \left[-0.5,0.5\right)$. \label{rho_x_collapse}}
\end{figure}

For a fixed mean density $\bar{\r}$, the model's exclusion interactions and diffusive dynamics yield a density profile that is a scaling function of $\el/L$, asserting that the density substantially deviates from $\bar{\r}$ throughout the \textit{entire} system. This stands in stark contrast to the non-interacting picture, where the width of the profile is $\sim\sqrt{D/r}$, with $D$ denoting the diffusion coefficient \cite{Search2}. Here we find that, for large systems, the density profile is entirely independent of the resetting rate $r$. In fact, it only depends on the mean density $\bar{\r}$, as seen in Eqs. \eqref{eq:rho_bulk_sol_3}-\eqref{eq:rho_bulk_sol_4}. A different picture emerges when the number of particles $N=\bar{\r}L$ is instead kept \textit{fixed}, in which case the density profile is shown to be a function of $\el$ alone, as obtained in Eq. \eqref{eq:rho_bulk_sol_2}. These MF predictions are verified, to remarkable precision, in data collapses obtained from extensive numerical simulations. Figure \ref{rho_x_collapse} demonstrates the density profile's scaling form for a fixed mean density while Fig. \ref{rho_ell_collapse} shows its behavior for fixed particle number $N$. Additional data is provided in the supplemental material (SM) for different parameters.

\begin{figure}
\begin{centering}
\includegraphics[scale=0.6]{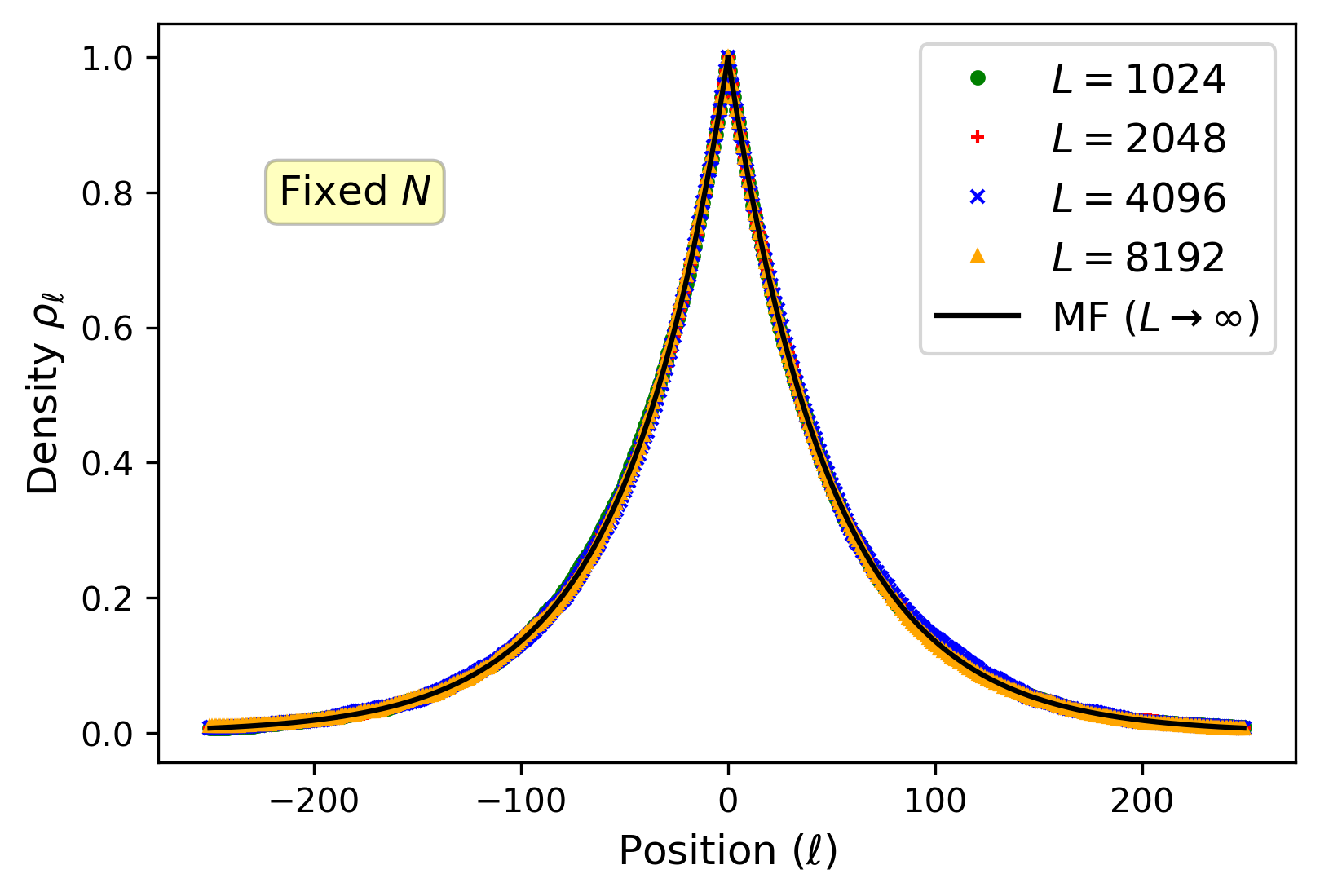}
\par\end{centering}
\caption{Data collapse of the density profile for $r=1$ and $\overline{\protect\r}L=N=100$
plotted versus the lattice index $\protect\el$ for 250 sites to the
left and right of the origin. The different markers denote various
values of the system size $L$ while the solid black curve denotes
the theoretical MF prediction in Eq. \eqref{eq:rho_bulk_sol_2} for
fixed $N$. Note that the site indices $\el$ have been shifted to $\el=-L/2+1,...,0,...,L/2$ for convenience of presentation. \label{rho_ell_collapse}}
\end{figure}

\textit{MF analysis} - To evoke the MF approximation, we first average over the Markov
chain in Eqs. \eqref{eq:markov chain} and \eqref{eq:markov Gamma},
replacing the mean occupation $\left\langle \t_{\el}\right\rangle $
by the corresponding density field $\r_{\el}\in\left[0,1\right]$. The MF approximation is then manifested in the factorization of products of the form $\left\langle \t_{\el}\t_{k}\right\rangle \approx\left\langle \t_{\el}\right\rangle \left\langle \t_{k}\right\rangle \ra\r_{\el}\r_{k}$.
The rates $R_{\el=0}$ and $R_{\el\ne0}$ in Eq. \eqref{eq:Gamma_sigma_R}
respectively become $\left\langle R_{\el=0}\right\rangle=r\sum_{m=1}^{L-1}\r_{m}$ and $\left\langle R_{\el\ne 0}\right\rangle =r\r_{\el}$.
While it is generally quite difficult to rigorously justify the MF approximation, this approach has proven to be remarkably successful for analyzing numerously many lattice models with exclusion interactions \cite{burlatsky1992, burlatsky1996, De_Coninck1997, Tsekouras2008, Cividini2017, Miron_2020, miron2020driven}. As we show below, MF also provides a remarkably accurate description of the present model.

In the limit $dt\ra0$, the MF approximation yields the following
equation for $\r_{\el}\left(t\right)$, 
\begin{equation}
\partial_{t}\r_{\el}=\r_{\el+1}-2\r_{\el}+\r_{\el-1}+\s_{\el}\left(1-\r_{0}\right)\langle R_{\el}\rangle.\label{eq:density eqn}
\end{equation}
Our interest lies in the stationary behavior of the density profile. To this end, we set $\partial_{t}\r_{\el}=0$
and \textit{separately} analyze Eq. \eqref{eq:density eqn} at sites
$\el\ne0$, termed the ``bulk'' equation, and at site $\el=0$,
which we call the ``boundary'' equation. Solving the stationary
bulk equation gives
\begin{equation}
\r_{\el}=c_{1}A_{-}^{\el}+c_{2}A_{+}^{\el},\label{eq:rho_bulk_sol_1}
\end{equation}
where $c_{1,2}$ are constants and we have defined
\begin{equation}
A_{\pm}=1+\frac{a}{2}\left(1\pm\sqrt{1+\frac{4}{a}}\right),\label{eq:A_pm}
\end{equation}
and
\begin{equation}
a=r\left(1-\r_{0}\right).\label{eq:a}
\end{equation}
Using the dynamic's symmetry around site $\el=0$, i.e. $\r_{\el}=\r_{L-\el}$,
gives $c_{2}=c_{1}A_{+}^{-L}$. The particle conservation condition
\begin{equation}
N=\overline{\r}L=\r_{0}+\sum_{\el=1}^{L-1}\r_{\el},\label{eq:particle_conservation}
\end{equation}
is then used to determine $c_{1}$, such that $\r_{\el}$ becomes 
\begin{equation}
\r_{\el}=\frac{\left(1-A_{-}\right)\left(L\overline{\r}+a/r-1\right)}{2\left(A_{-}-A_{-}^{L}\right)}\left(A_{-}^{\el}+A_{+}^{\el-L}\right).\label{eq:rho_bulk_sol_2}
\end{equation}

Although we have obtained a formal solution for the MF density profile $\r_{\el}$, our work is not yet done since $\r_{\el}$ still depends on the density at site $\el=0$ through $a$ in Eq. \eqref{eq:a}. To determine $\r_{0}$, we revisit Eq. \eqref{eq:density eqn} for the density profile and consider its behavior at $\el=0$. In the stationary limit, this equation can be written as 
\begin{equation}
\r_{1}=\r_{0}-\frac{r}{2}\left(1-\r_{0}\right)\left(N-\r _0\right),\label{eq:rho_1_eqn}
\end{equation}
where $N-\r_0=\sum_{m=1}^{L-1}\r_{m}$ and we have again used the $\el\ra L-\el$ symmetry to replace $\r_{1}+\r_{L-1}\ra2\r_{1}$. To make progress, we separately consider two distinctly different physical scenarios: the case of a constant particle density $\bar{\r}$ and the case of a constant particle number $N$.

\textit{Fixed density $\overline{\r}$} - In this case the number
of particles in the system $N$ grows linearly with system-size $L$.
For large $L$, and a correspondingly large number of particles, there is a small time gap between the instance site $\el=0$ is vacated by a particle hopping to a vacant neighboring site, and the time it is reoccupied due to a resetting event. Reoccupation due to hopping from neighboring sites is negligible, since it is attempted with rate $1$ while resetting events are attempted with rate $L\bar{\r }r\equiv Nr\gg1$. In the limit of $L\ra\infty$, where the system contains infinitely many particles, the total rate of resetting attempts becomes infinite and reoccupation is \textit{immediate}. We thus expect
that the density near the origin be unity, up to small finite-$L$ corrections.
We thus consider the ansatz 
\begin{equation}
\begin{cases}
\r_{0}\asy1-(\a/L)^{\m}\\
\r_{1}\asy1-(\b/L)^{\n}
\end{cases},\label{eq:ansatz}
\end{equation}
generally allowing for a different scaling with $L$ at sites $\el=0$ and $\el\ne0$. 
Substituting this ansatz into Eq. \eqref{eq:rho_1_eqn} yields the relation $\m=1+\n$. We can then determine the value of $\n$ by substituting Eq. \eqref{eq:ansatz} into Eq. \eqref{eq:rho_bulk_sol_2} for $\r_\el$, evaluated at site $\el=1$. This self-consistency requirement sets $\n=1$ and, correspondingly, $\m=2$ (see the SM for details). With this, we obtain the density profile at sites $\el\ne0$ as
\begin{equation}
\r_{\el}\asy 2^{-1}\a\overline{\r}\left(e^{\a}-1\right)^{-1}\left(e^{\a\left(L-\el\right)/L}+e^{\a\el/L}\right).\label{eq:rho_bulk_sol_3}
\end{equation}
The parameter $\a$ is determined by demanding that the ansatz for $\r_0$ in Eq. \eqref{eq:ansatz} be consistent with the density profile $\r_\el$ in Eq. \eqref{eq:rho_bulk_sol_3} at site $\el=0$. This gives 
\begin{equation}
\overline{\r}\a\coth\left[\a/2\right]=2,\label{eq:alpha_equation}
\end{equation}
providing  a transcendental equation for $\a$ that we must numerically solve, given the value of $\overline{\r}$. With this, the density profile assumes the simpler form
\begin{equation}
\r_{\el}\asy \cosh{\left[\frac{\a}{2}\left(\frac{L-2\el}{L}\right)\right]}/\cosh{\left[\frac{\a}{2}\right]}.\label{eq:rho_bulk_sol_4}
\end{equation}

\begin{figure}[t]
\begin{centering}
\includegraphics[scale=0.6]{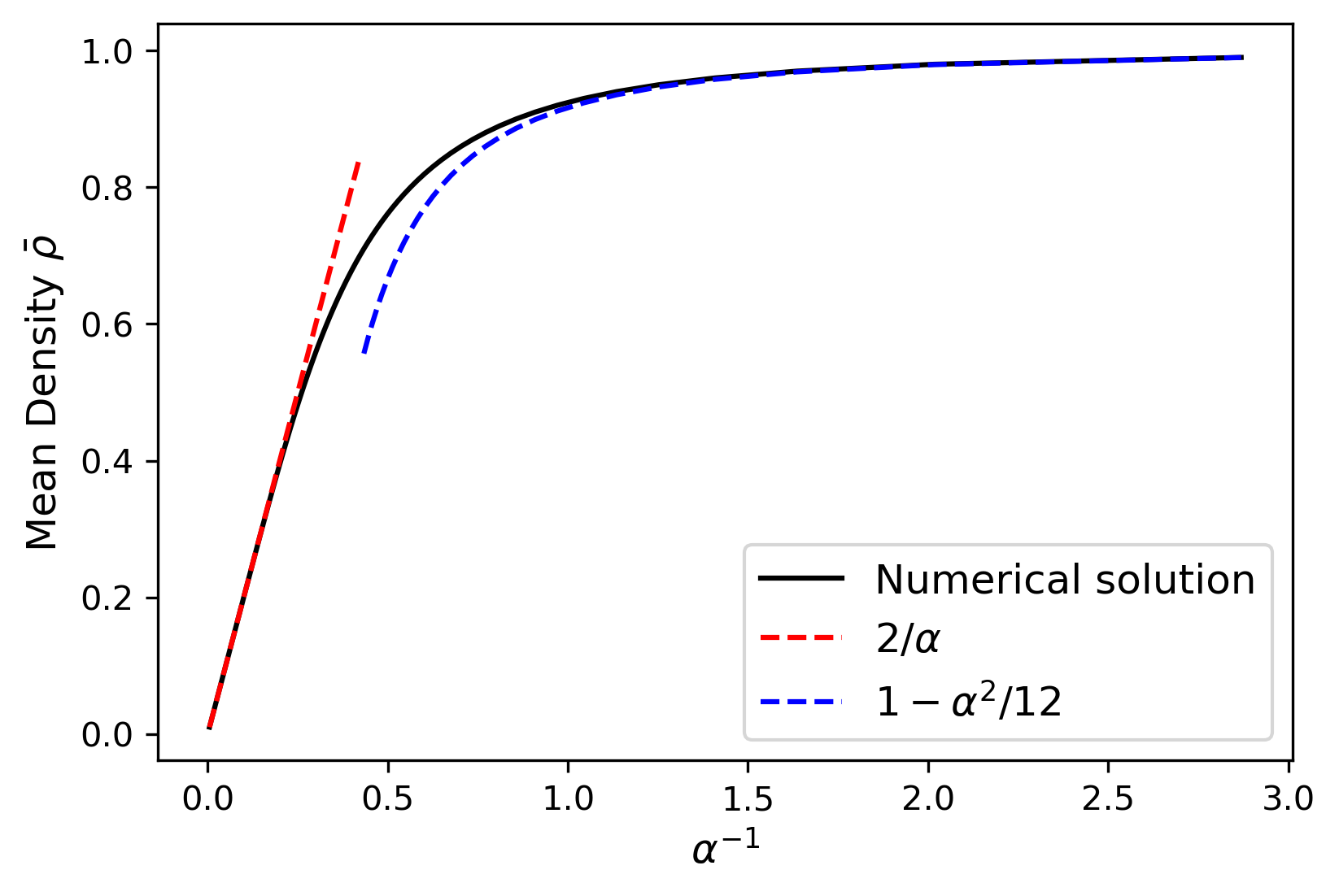}
\par\end{centering}
\caption{Plot of $\bar{\rho}$ versus $\alpha$, as obtained from Eq. \eqref{eq:alpha_equation} for $r=1$. The solid black curve is the numerical solution of the equation, the dashed red curve is the large $\alpha$ approximation and the dashed blue curve describes the small $\alpha$ limit.\label{rho_alpha_plot}}
\end{figure}

Although the relation between $\a$ and $\bar{\r}$ cannot be inverted analytically, we can still obtain important insight regarding the dependence of $\alpha$ on $\bar{\r}$ in the asymptotic limits of $\a\ra0$ and $\a\ra\infty$. A straightforward analysis yields
\begin{equation}
\overline{\r}\asy\begin{cases}
1-\a^{2}/12 & \a\ll1\\
2/\a & \a\gg1
\end{cases},\label{eq:rho_alpha_relation}
\end{equation}
as demonstrated in Fig. \ref{rho_alpha_plot}.
We conclude that for a fixed mean density $\overline{\r}$, $\r_\el$ becomes a scaling function of $\el/L$, at large $L$. This implies that the density profile spans the entire system, as is clearly observed in Fig. \ref{rho_x_collapse}. This scaling behavior also appears for $r\pro L$, as shown in the SM. 

\textit{Fixed particle number $N$} - We next consider the case where the particle number $N$ is fixed, finite, and independent of $L$. The time gap between the instance site $\el=0$ is vacated and the instance it is reoccupied now remains \textit{finite}, even as $L$ is increased. Correspondingly, both $\r_{0}$ and $\r_{1}$ are asymptotically independent of $L$. With this insight, we return to Eq. \eqref{eq:rho_bulk_sol_2} in order to relate $\r_{\el}$ at $\el=0$ to $\r_{0}$ for finite $N$, in the limit of large $L$. Substituting $\overline{\r}=N/L$ and $a=r\left(1-\r_{0}\right)$ into Eq. \eqref{eq:rho_bulk_sol_2}, setting $\el=0$ and equating to $\r_{0}$ asymptotically yields the polynomial equation 
\begin{equation}
\frac{r\left(1-\r_{0}\right)\left(N-\r_{0}\right)}{\sqrt{r\left(1-\r_{0}\right)\left(r\left(1-\r_{0}\right)+4\right)}-r\left(1-\r_{0}\right)}\asy\r_{0},\label{eq:rho_0_eqn}
\end{equation}
where we have used the fact that $A_{+}^{-L}$ and $A_{-}^{L}$ both vanish as $L\ra\infty$, for any $L$-independent $a$ (see Eq. \eqref{eq:A_pm}).
This equation for $\r_{0}$ has three roots, of which only one satisfies $\r_{0}\in\left[0,1\right]$. While its precise analytical form is rather involved and is thus deferred to the SM, in the limit $1\ll N\ll L$, where the system contains many particles but the average density is low, it reduces to $\r_{0}= 1-4r^{-1}N^{-2}+\ord{N^{-4}}$. By  Eq. \eqref{eq:a}, this translates into $a\asy4/N^2$ which, when substituted into Eq. \eqref{eq:rho_bulk_sol_2}, reveals that the density profile has a width of $\sim N/2$ around site $\el=0$.  Since $N$ is fixed, the density profile remains a function of $\el$ alone, implying that it only extends over a finite $\sim\ord N$ region near the origin, as demonstrated in Fig. \ref{rho_ell_collapse}.

\textit{Conclusions} - In this Letter, we have gone beyond the popular renewal framework to study the effect of \textit{local} resetting on a many-body interacting system. Employing a MF approach, we derived the stationary density profile of $N$ particles diffusing on a 1D ring lattice of $L$ sites while being subject to exclusion interactions and local resetting with rate $r$. For large systems occupied by many particles, we find that the profile's width is independent of $r$, being solely a function of the position and average density $\bar{\r}$ (or the number of particles $N$, depending on which was kept fixed while taking the large system limit). This intriguing behavior directly follows from the delicate interplay between local resetting and exclusion interactions, which prohibit resetting if the origin is already occupied. Indeed, for non-interacting particles the density profile is known to adopt a width $\sim r^{-1/2}$ \cite{Search2}, irrespective of $\bar{\r}$ or $N$, which stands in stark contrast to our findings here.  

Two recent experimental studies explored resetting in the context of diffusive single-particle systems, clearly marking interacting many-body systems as the next frontier \cite{Roichman2020,Ciliberto2020}. Exclusion is one complication that is sure to arise in such systems, and the results established herein are thus especially well-posed to serve as a benchmark for comparison. Moreover, the approach we presented can be extended further, as required, to capture more realistic scenarios featuring excluded volume interactions and local resetting. These will be considered elsewhere. 

\textit{Acknowledgments} - A. M. wishes to thank David Mukamel for his ongoing encouragement and support. A. M. also thanks the support of the Center of Scientific Excellence at the Weizmann Institute of Science. S. R. acknowledges support from the Azrieli Foundation, from the Raymond and Beverly Sackler Center for Computational Molecular and Materials Science at Tel Aviv University, and from the Israel Science Foundation (grant No. 394/19).

\setcounter{figure}{0}
\renewcommand{\thefigure}{S\arabic{figure}}

\newpage
\part*{Supplemental Material}

\section{Additional comparison between simulations and mean-field predictions}

In this section we further establish the effectiveness of the mean-field (MF) approach in the study of the model presented in the main text. We do so by extending the comparison provided in the main text, between the MF result for the density profile $\r_{\el}$ of the main text Eq. (14) and direct numerical simulations of the model, for additional parameters. 

\begin{figure}
\begin{centering}
\includegraphics[scale=0.6]{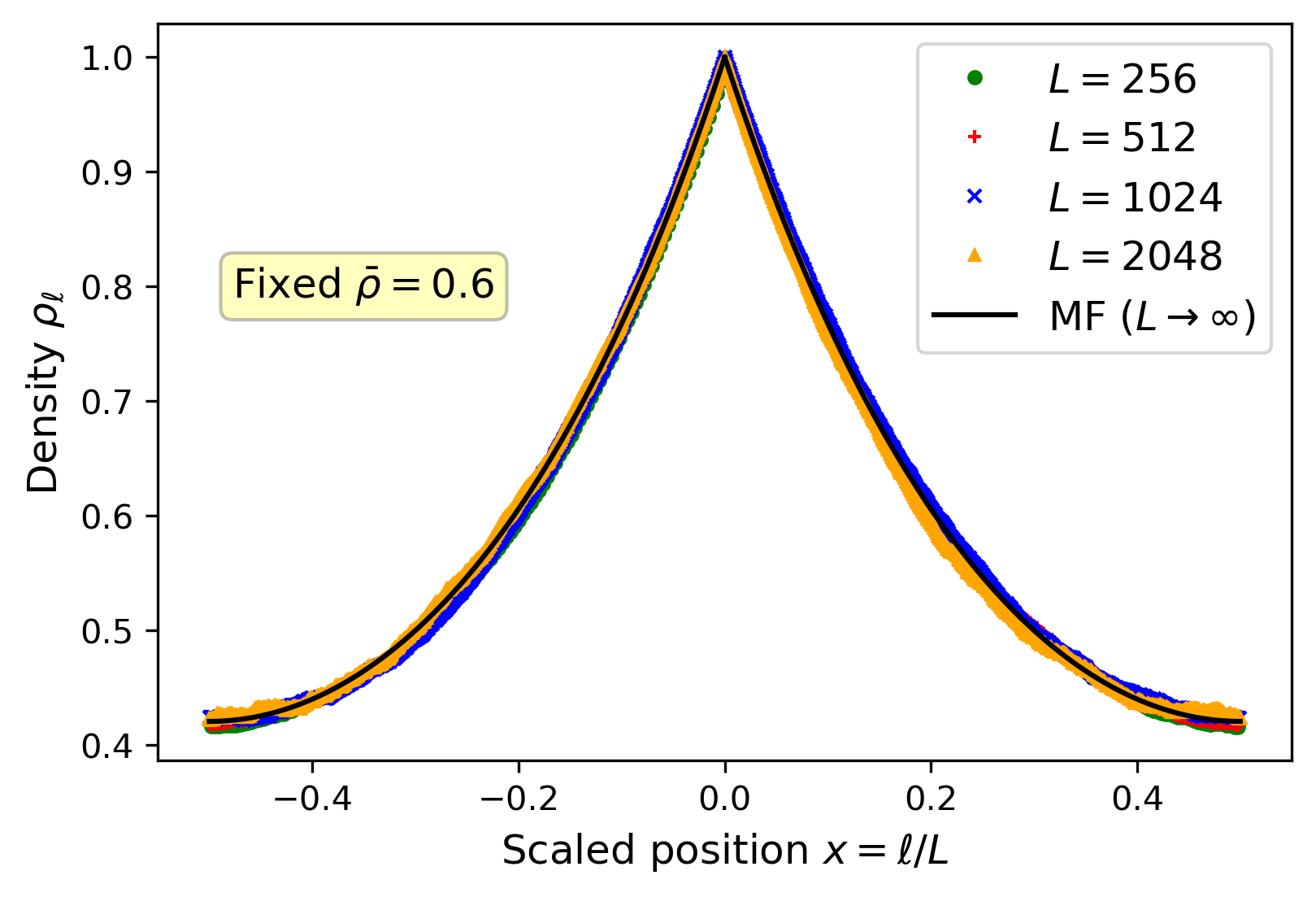}
\par\end{centering}
\caption{Data collapse of the density profile versus the scaling variable $x=\protect\el/L$
for $r=1$ and $\bar{\r}=0.6$. The different markers denote the
four values of $L$ and
the solid black curve denotes the numerical solution to the main text
mean-field Eq. $\left(14\right)$.}
\label{collapse1}
\end{figure}

\begin{figure}
\begin{centering}
\includegraphics[scale=0.6]{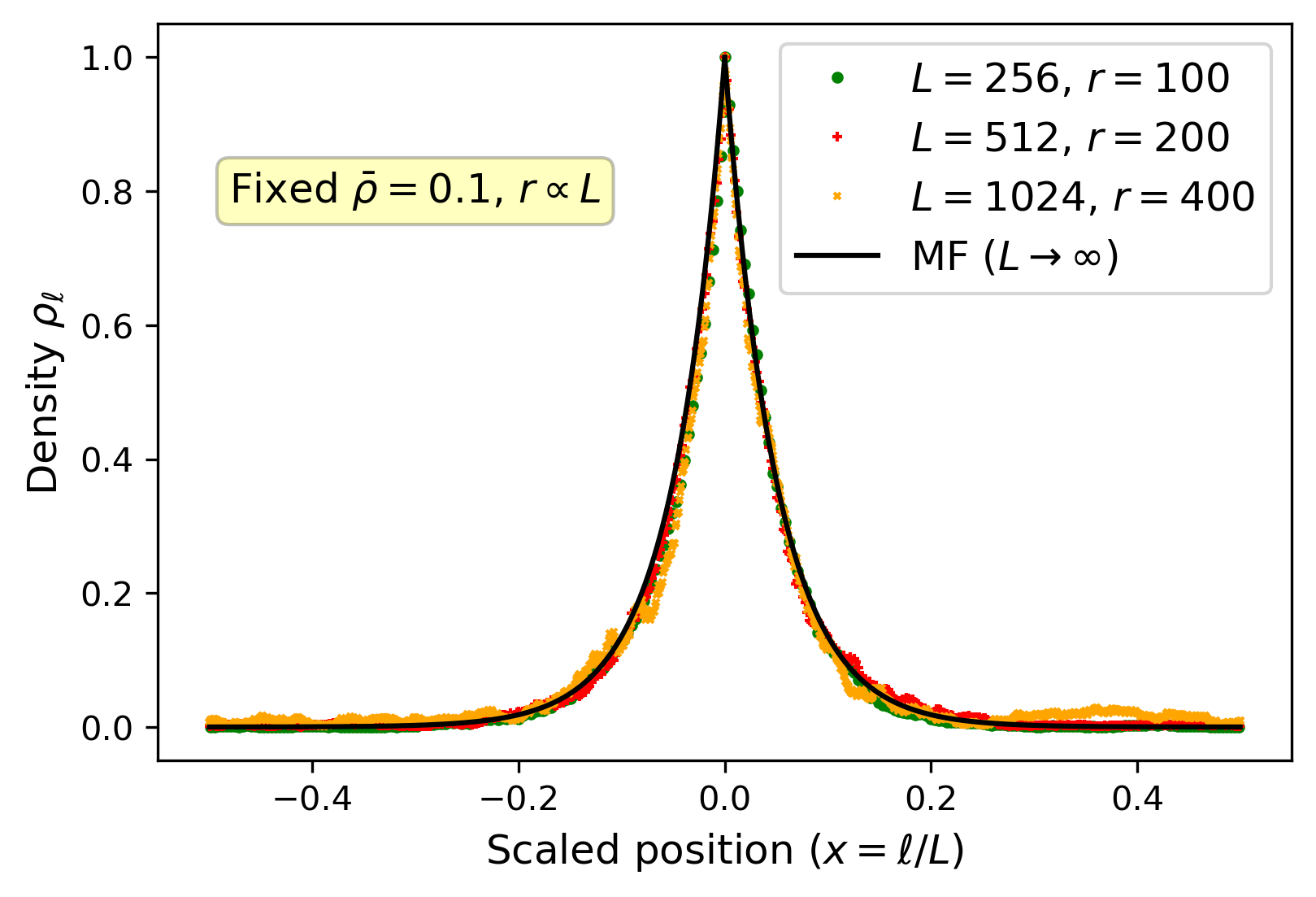}
\par\end{centering}
\caption{Data collapse of the density profile versus the scaling variable $x=\protect\el/L$
for $\overline{\protect\r}=0.1$. The different markers denote the
three values of $L$ and corresponding $r\protect\pro L$, whereas
the solid black curve denotes the numerical solution to the main text
mean-field Eq. $\left(12\right)$.}
\label{collapse}
\end{figure}

The main text Fig. 2 compares a data collapse (for different values of the system size $L$) of the density profile versus $x=\el/L$ to our MF results, for a resetting rate $r=1$ and a mean density $\bar{\r}=0.2$. Figure \ref{collapse1} complements this by providing the same comparison for the mean density $\bar{\r}=0.6$ and the resetting rate $r=1$. In Fig. \ref{collapse} we provide an additional comparison for a resetting rate which grows linearly with $L$ and a mean density $\overline{\r}=0.1$. The remarkable agreement between MF theory and numerical simulations of the density profile versus $x=\el/L$ reassures that MF provides a very good description for a broad range of model parameters and implies that the profile indeed remains a scaling function of $\el/L$, even when $r\pro L$.

\section{System-size scaling of $\r_\el$}
Here we set the values of $\n$ and $\m$ in the ansatz $\r_{0}=1-\left(\frac{\a}{L}\right)^{\m}$ and $\r_{1}=1-\left(\frac{\b}{L}\right)^{\n}$ that appear in the main text Eq. (11). To this end, we first substitute them into the stationary boundary equation $\r_{1}=\r_{0}-\frac{r}{2}\left(1-\r_{0}\right)\left(L\overline{\r}-\r_{0}\right)$ of the main text Eq. (10).
We obtain
\begin{equation}
\left(\frac{\b}{L}\right)^{\n}-\frac{r\overline{\r}\a^{\m}}{2L^{\m-1}}-\frac{r}{2}\left(\frac{\a}{L}\right)^{2\m}-\left(1-\frac{r}{2}\right)\left(\frac{\a}{L}\right)^{\m}=0.\label{eq:substitution}
\end{equation}
As $L\ra\infty$, this can only be satisfied if $\m>1$, in which case Eq. \eqref{eq:substitution} becomes
\begin{equation}
\frac{r\overline{\r}\a^{\m}}{2L^{\m-1}}\asy\left(\frac{\b}{L}\right)^{\n}.\label{eq:substitution_lim}
\end{equation}
Correspondingly, we conclude
\begin{equation}
\begin{cases}
\m=1+\n\\
\b^{\n}\asy\frac{r\overline{\r}}{2}\a^{1+\n}
\end{cases}.\label{eq:beta_gamma}
\end{equation}

\begin{figure}[H]
\begin{centering}
\includegraphics[scale=0.6]{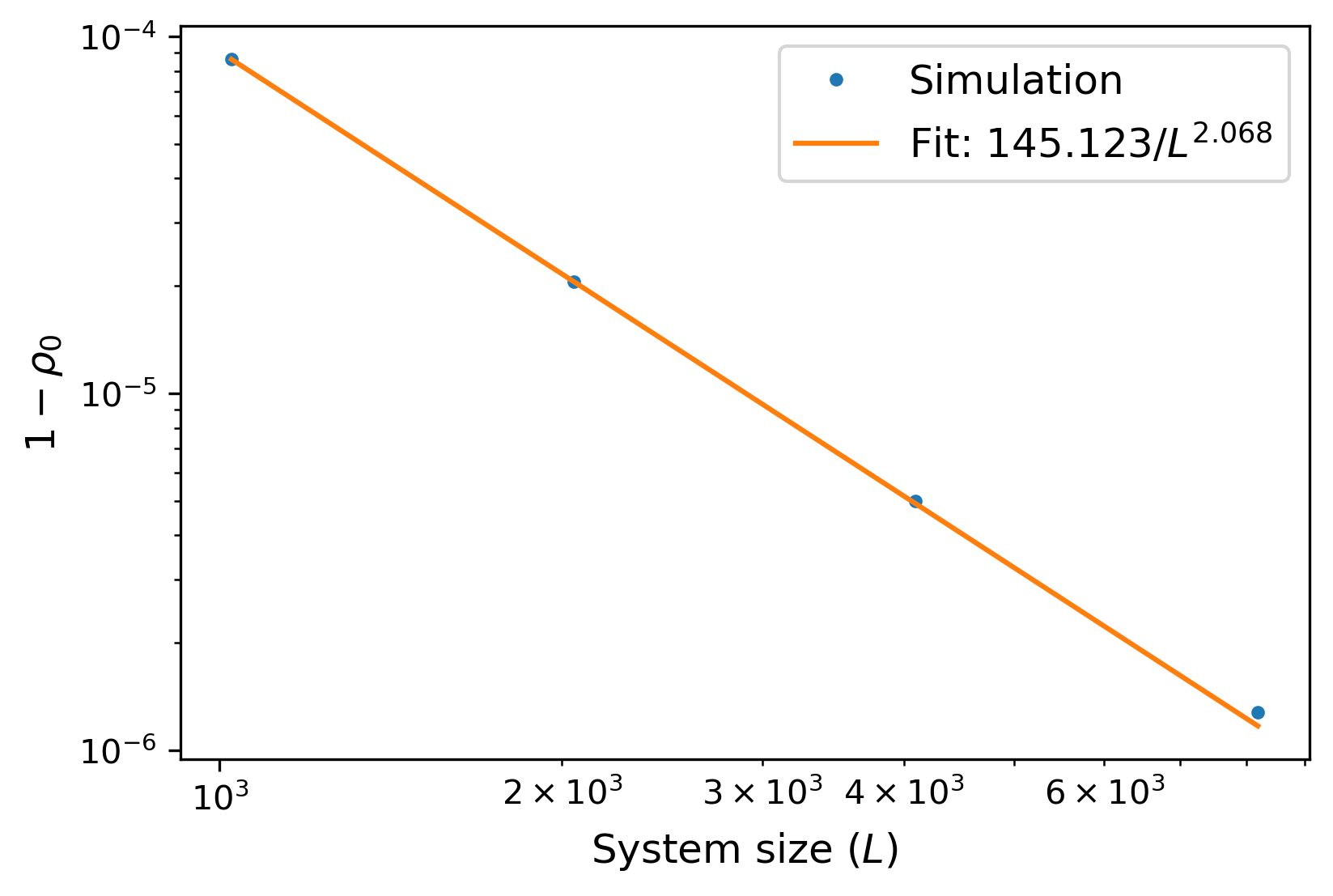}
\par\end{centering}
\caption{Log-log plot of $\left(1-\protect\r_{0}\right)$ versus $L$ for $r=1$ and $\overline{\protect\r}=0.2$. The blue dots denote the model's numerical simulation results while the solid orange line a least-squares fit to $c_{0}/L^{c_{1}}$.}
\label{1_min_rho_0_loglog}
\end{figure}

Having related $\m$ to $\n$, we next set out to derive the value of $\n$. To do this, we substitute the ansatz at site $\el=0$, i.e. $\r_{0}=1-\left(\frac{\a}{L}\right)^{1+\n}$,
into the main text Eq. (9) for the density profile $\r_{\el}$ at site $\el=1$
\begin{equation}
\r_{1}=\frac{\left(1-A_{-}\right)\left(L\overline{\r}+\frac{a}{r}-1\right)}{2\left(A_{-}-A_{-}^{L}\right)}\left(A_{-}+A_{+}^{1-L}\right),\label{eq:rho_1}
\end{equation}
and demand that it identifies with our ansatz, $\r_{1}=1-\left(\frac{\b}{L}\right)^{\n}$.
Recall that $A_{\pm}$ is provided in the main text Eq. (6) as $A_{\pm}=1+\frac{a}{2}\left(1\pm\sqrt{1+\frac{4}{a}}\right)$ and that $a=r\left(1-\r_{0}\right)$. Using the ansatz for $\r_0$, we can rewrite $A_{\pm}^{L}$ in the limit of large $L$ as
\[
A_{\pm}^{L}\asy\left[1+\frac{r}{2}\left(\frac{\alpha}{L}\right)^{\n+1}\left(1\pm\sqrt{\frac{4}{r}\left(\frac{L}{\alpha}\right)^{\n+1}}\right)\right]^{L}
\]
\begin{equation}
\asy\left[1\pm\sqrt{r}\left(\frac{\alpha}{L}\right)^{\frac{\n+1}{2}}\right]^{L}\asy e^{\pm\sqrt{r}\a^{\frac{\n+1}{2}}L^{\frac{1-\n}{2}}}.\label{eq:A_to_the_L}
\end{equation}
Depending on the value of $\n$, $A_{\pm}^{L}$ exhibits one of three distinct behaviors as $L\ra\infty$: For $0<\n<1$, it exponentially decays/blows-up, while for $\n>1$ we get $A_{\pm}^{L}\asy1\pm\ord{L^{\frac{1-\n}{2}}}.$ Yet, for $\n=1$, we obtain the $L$-independent expression $A_{\pm}^{L}\asy e^{\pm\sqrt{r}\a}$.
We finally use this understanding to derive the value of $\n$. It is straightforward to show that the leading, large-$L$ behavior of $\r_1$ in Eq. \eqref{eq:rho_1} is 
\begin{equation}
\r_1\asy L^{\frac{1-\n}{2}}.\label{eq:rho_1_asy}
\end{equation}
This can only self-consistently agree with the ansatz $\r_{1}=1-\left(\frac{\b}{L}\right)^{\n}$ as $L\ra\infty$ if $\n=1$, which immediately also allows us to deduce $\m=2$.
The large-$L$ scaling of the deviation of the density $\r_0$ from unity is numerically verified in Fig. \ref{1_min_rho_0_loglog}.

\section{Fixed number of particles $N$}

In the case of a fixed number of particles $N$, the polynomial main text Eq. (16) is solved by
\begin{equation}
\r_{0}=\frac{4+r-\left(-1\right)^{1/3}\chi^{1/3}\left(1+\frac{\left(1+3N^{2}\right)r^{2}+8r+16}{\left(-1\right)^{2/3}\chi^{2/3}}\right)}{3r},\label{eq:rho_0_SM}
\end{equation}
where 
\[
\c\left(r,N\right)=\left(1-9N^{2}\right)r^{3}+6\left(2+3N^{2}\right)r^{2}
\]
\[
+48r+64+3iN\sqrt{3r^{3}}
\]
\begin{equation}
\times\sqrt{\left(1-N^{2}\right)^{2}r^{3}+4\left(3+5N^{2}\right)r^{2}+4\left(12+N^{2}\right)r+64}.\label{eq:chi_SM}
\end{equation}
Although it is far from obvious at first sight, $\r_{0}$ in Eqs. \eqref{eq:rho_0_SM} and \eqref{eq:chi_SM} satisfies the requirement $\r_{0}\in\left[0,1\right]$ for $r,N>0$.

\begin{thebibliography}{1}

\bibitem{Search1}
Eliazar, I., Koren, T. and Klafter, J., 2007. Searching circular DNA strands. Journal of Physics: Condensed Matter, 19(6), p.065140.

\bibitem{Search2}
Evans, M.R. and Majumdar, S.N., 2011. Diffusion with stochastic resetting. Physical review letters, 106(16), p.160601.

\bibitem{Search3}
Evans, M.R., Majumdar, S.N. and Mallick, K., 2013. Optimal diffusive search: nonequilibrium resetting versus equilibrium dynamics. Journal of Physics A: Mathematical and Theoretical, 46(18), p.185001.

\bibitem{Search4}
Kusmierz, L., Majumdar, S.N., Sabhapandit, S. and Schehr, G., 2014. First order transition for the optimal search time of Levy flights with resetting. Physical review letters, 113(22), p.220602.

\bibitem{Search5} 
Bhat, U., De Bacco, C. and Redner, S., 2016. Stochastic search with Poisson and deterministic resetting. Journal of Statistical Mechanics: Theory and Experiment, 2016(8), p.083401.

\bibitem{Search6} 
Pal, A. and Reuveni, S., 2017. First Passage under Restart. Physical review letters, 118(3), p.030603.

\bibitem{Search7} 
Chechkin, A. and Sokolov, I.M., 2018. Random search with resetting: a unified renewal approach. Physical review letters, 121(5), p.050601.

\bibitem{Search8}
Kusmierz, L. and Toyoizumi, T., 2019. Robust random search with scale-free stochastic resetting. Physical Review E, 100(3), p.032110.

\bibitem{Search9} 
Pal, A., Kusmierz, L. and Reuveni, S., 2020. Search with home returns provides advantage under high uncertainty. Phys. Rev. Research 2, 043174.

\bibitem{Search10} 
Bressloff, P.C., 2020. Directed intermittent search with stochastic resetting. Journal of Physics A: Mathematical and Theoretical, 53(10), p.105001.

\bibitem{Luby}
Luby, M., Sinclair, A., and Zuckerman, D., 1993. Optimal speedup of Las Vegas algorithms. Inf. Process Lett. 47, 173.

\bibitem{Gomes}
Gomes, C. P., Selman, B., and Kautz, H. 1998. Boosting combinatorial search through randomization, AAAI/IAAI 98, 431.

\bibitem{Steiger}
Steiger, D.S., Ronnow, T.F. and Troyer, M., 2015. Heavy tails in the distribution of time to solution for classical and quantum annealing. Physical review letters, 115(23), p.230501.

\bibitem{Optimization1}
Montanari, A., and Zecchina, R. 2002. Optimizing searches via rare events, Phys.
Rev. Lett. 88, 178701.

\bibitem{Optimization2}
T. Rotbart, S. Reuveni, and M. Urbakh, Michaelis-Menten reaction scheme as a unified approach towards the optimal restart problem, Phys. Rev. E 92,
060101 (2015).

\bibitem{Optimization3}
Belan, S., 2018. Restart could optimize the probability of success in a Bernoulli trial. Physical review letters, 120(8), p.080601.

\bibitem{Optimization4}
De Bruyne, B., RandonFurling, J. and Redner, S., 2020. Optimization in First-Passage Resetting. Physical Review Letters, 125(5), p.050602.

\bibitem{Optimization5}
De Bruyne, B., Randon-Furling, J. and Redner, S., 2020. Optimization and Growth in First-Passage Resetting. arXiv preprint arXiv:2009.03419.

\bibitem{Reuveni}
Reuveni, S., Urbakh, M., and Klafter, J., 2014. Role of substrate unbinding in Michaelis-Menten enzymatic reactions, Proc. Natl. Acad. Sci. U. S. A. 111, 4391.

\bibitem{Berezhkovskii}
A. M. Berezhkovskii, A. Szabo, T. Rotbart, M. Urbakh, and A. B. Kolomeisky, Dependence of the Enzymatic Velocity on the Substrate Dissociation Rate, J. Phys. Chem. B 121, 3437 (2016).

\bibitem{Robin}
T. Robin, S. Reuveni, and M. Urbakh, Single-molecule theory of enzymatic inhibition, Nat. Commun. 9, 779 (2018).

\bibitem{Lapeyre}
Lapeyre, G.J. and Dentz, M., 2017. Reaction-diffusion with stochastic decay rates. Physical Chemistry Chemical Physics, 19(29), pp.18863-18879.

\bibitem{Roldan1} Roldan, E., Lisica, A., Sanchez-Taltavull, D. and Grill, S.W., 2016. Stochastic resetting in backtrack recovery by RNA polymerases. Physical Review E, 93(6), p.062411.

\bibitem{Budnar} Budnar, S., Husain, K.B., Gomez, G.A., Naghibosadat, M., Varma, A., Verma, S., Hamilton, N.A., Morris, R.G. and Yap, A.S., 2019. Anillin promotes cell contractility by cyclic resetting of RhoA residence kinetics. Developmental cell, 49(6), pp.894-906.

\bibitem{renewalf1} 
Evans, M.R., Majumdar, S.N. and Schehr, G., 2020. Stochastic resetting and applications. Journal of Physics A: Mathematical and Theoretical.

\bibitem{renewal1} 
Meylahn, J.M., Sabhapandit, S. and Touchette, H., 2015. Large deviations for Markov processes with resetting. Physical Review E, 92(6), p.062148.

\bibitem{renewal2}
Evans, M.R. and Majumdar, S.N., 2018. Run and tumble particle under resetting: a renewal approach. Journal of Physics A: Mathematical and Theoretical, 51(47), p.475003.

\bibitem{renewal3} 
Evans, M.R. and Majumdar, S.N., 2018. Effects of refractory period on stochastic resetting. Journal of Physics A: Mathematical and Theoretical, 52(1), p.01LT01.

\bibitem{renewal4}
Chatterjee, A., Christou, C. and Schadschneider, A., 2018. Diffusion with resetting inside a circle. Physical Review E, 97(6), p.062106.

\bibitem{renewal5}
Kusmierz, L. and Gudowska-Nowak, E., 2019. Subdiffusive continuous-time random walks with stochastic resetting. Physical Review E, 99(5), p.052116.

\bibitem{renewal6}
Maso-Puigdellosas, A., Campos, D. and Mendez, V., 2019. Transport properties of random walks under stochastic noninstantaneous resetting. Physical Review E, 100(4), p.042104.

\bibitem{renewal7}
Den Hollander, F., Majumdar, S.N., Meylahn, J.M. and Touchette, H., 2019. Properties of additive functionals of Brownian motion with resetting. Journal of Physics A: Mathematical and Theoretical, 52(17), p.175001.

\bibitem{renewal8}
Ahmad, S., Nayak, I., Bansal, A., Nandi, A. and Das, D., 2019. First passage of a particle in a potential under stochastic resetting: A vanishing transition of optimal resetting rate. Physical Review E, 99(2), p.022130.

\bibitem{renewal9}
Bodrova, A.S., Chechkin, A.V. and Sokolov, I.M., 2019. Scaled Brownian motion with renewal resetting. Physical Review E, 100(1), p.012120.

\bibitem{renewal10}
Ray, S., Mondal, D. and Reuveni, S., 2019. Peclet number governs transition to acceleratory restart in drift-diffusion. Journal of Physics A: Mathematical and Theoretical, 52(25), p.255002.

\bibitem{renewal11}
Pal, A., Chatterjee, R., Reuveni, S., and Kundu, A., 2019. Local time of diffusion with stochastic resetting. J. Phys. A. 52, 264002.

\bibitem{renewal12}
Masoliver, J., 2019. Telegraphic processes with stochastic resetting. Physical Review E, 99(1), p.012121.

\bibitem{renewal13}
Pal, A. and Prasad, V.V., 2019. First passage under stochastic resetting in an interval. Physical Review E, 99(3), p.032123.

\bibitem{renewal14} Pal, A., Kusmierz, L. and Reuveni, S., 2019. Time-dependent density of diffusion with stochastic resetting is invariant to return speed. Physical Review E, 100(4), p.040101.

\bibitem{renewal15}
Gupta, D., Plata, C.A. and Pal, A., 2020. Work fluctuations and Jarzynski equality in stochastic resetting. Physical Review Letters, 124(11), p.110608.

\bibitem{renewal16}
Ray, S. and Reuveni, S., 2020. Diffusion with resetting in a logarithmic potential, J. Chem. Phys. 152, 234110.

\bibitem{renewal17}
Bodrova, A.S. and Sokolov, I.M., 2020. Resetting processes with noninstantaneous return. Physical Review E, 101(5), p.052130.

\bibitem{renewal18}
Gupta, D., Plata, C.A. and Pal, A., 2020. Work fluctuations and Jarzynski equality in stochastic resetting. Physical Review Letters, 124(11), p.110608.

\bibitem{renewal19}
Belan, S., 2020. Median and mode in first passage under restart. Physical Review Research, 2(1), p.013243.

\bibitem{renewal20}
Singh, R.K., Metzler, R. and Sandev, T., 2020. Resetting dynamics in a confining potential. Journal of Physics A: Mathematical and Theoretical.

\bibitem{universal1} 
Reuveni, S., 2016. Optimal stochastic restart renders fluctuations in first passage times universal. Physical review letters, 116(17), p.170601.

\bibitem{universal2} 
Pal, A., Eliazar, I. and Reuveni, S., 2019. First passage under restart with branching. Physical review letters, 122(2), p.020602.

\bibitem{universal3} Pal, A. and Prasad, V.V., 2019. Landau-like expansion for phase transitions in stochastic resetting. Physical Review Research, 1(3), p.032001.

\bibitem{universal4} Pal, A., Kusmierz, L. and Reuveni, S., 2019. Invariants of motion with stochastic resetting and space-time coupled returns. New Journal of Physics, 21(11), p.113024.

\bibitem{universal5}
Eliazar, I. and Reuveni, S., 2020. Mean-performance of sharp restart I: Statistical roadmap. J. Phys. A: Math. Theor. 53 405004.

\bibitem{global1}Basu, U., Kundu, A. and Pal, A., 2019. Symmetric exclusion process under stochastic resetting. Physical Review E, 100(3), p.032136.

\bibitem{global2}Magoni M., Majumdar S. N. and Schehr G., 2020. Phys. Rev. Research 2, 033182.

\bibitem{global3}Karthika, S. and Nagar, A., 2020. Totally asymmetric simple exclusion process with resetting. Journal of Physics A: Mathematical and Theoretical, 53(11), p.115003.

\bibitem{Roichman2020}
Tal-Friedman, O., Pal, A., Sekhon, A., Reuveni, S. and Roichman, Y., 2020. Experimental realization of diffusion with stochastic resetting. J. Phys. Chem. Lett. 11, 7350-7355.

\bibitem{Ciliberto2020}
Besga, B., Bovon, A., Petrosyan, A., Majumdar, S. and Ciliberto, S., 2020. Optimal mean first-passage time for a Brownian searcher subjected to resetting: experimental and theoretical results. Phys. Rev. Research 2, 032029(R).

\bibitem{harris1965diffusion}
Theodore~E Harris.
\newblock Diffusion with collisions between particles.
\newblock {\em Journal of Applied Probability}, 2(2):323--338, 1965.

\bibitem{jepsen1965dynamics}
DW~Jepsen.
\newblock Dynamics of a simple many-body system of hard rods.
\newblock {\em Journal of Mathematical Physics}, 6(3):405--413, 1965.

\bibitem{percus1974anomalous}
Jerome~K Percus.
\newblock Anomalous self-diffusion for one-dimensional hard cores.
\newblock {\em Physical Review A}, 9(1):557, 1974.

\bibitem{alexander1978diffusion}
S~Alexander and P~Pincus.
\newblock Diffusion of labeled particles on one-dimensional chains.
\newblock {\em Physical Review B}, 18(4):2011, 1978.

\bibitem{burlatsky1992}
Burlatsky, S. F., et al. 
\newblock Directed walk in a one-dimensional lattice gas. 
\newblock {\em Physics Letters A}, 166.3-4 (1992): 230-234.

\bibitem{burlatsky1996}
Burlatsky, S. F., et al. 
\newblock Motion of a driven tracer particle in a one-dimensional symmetric lattice gas. \newblock {\em Physical Review E} 54.4 (1996): 3165.

\bibitem{De_Coninck1997}
De Coninck, J., G. Oshanin, and M. Moreau. \newblock Dynamics of a driven probe molecule in a liquid monolayer. 
\newblock {\em Europhysics Letters}, 38.7 (1997): 527.

\bibitem{Tsekouras2008}
Tsekouras, Konstantinos, and A. B. Kolomeisky. \newblock Parallel coupling of symmetric and asymmetric exclusion processes. 
\newblock {\em Journal of Physics A: Mathematical and Theoretical}, 41.46 (2008): 465001.

\bibitem{Cividini2017}
Cividini, Julien, David Mukamel, and H. A. Posch. 
\newblock Driven tracers in narrow channels. \newblock {\em Physical Review E}, 95.1 (2017): 012110.

\bibitem{Miron_2020}
Asaf Miron, David Mukamel, and Harald~A Posch.
\newblock Phase transition in a 1d driven tracer model.
\newblock {\em Journal of Statistical Mechanics: Theory and Experiment},
  2020(6):063216, jun 2020.

\bibitem{miron2020driven}
Asaf Miron and David Mukamel.
\newblock Driven tracer dynamics in a one dimensional quiescent bath.
\newblock {\em arXiv preprint arXiv:2007.08168}, 2020.

\end{thebibliography}
\end{document}